\documentclass[letter,twocolumn]{jpsj3}
\usepackage{txfonts}
\usepackage{graphicx}

\newcommand{\simj}{\stackrel{>}{_\sim}}
\newcommand{\simk}{\stackrel{<}{_\sim}}

\begin{document}
\title{Transition Temperature of Superconductivity in Sodium Tungsten Bronze \\
 - Theoretical Study Based on  First-principles Calculations - }

\author{Kazuhiro  {\sc Sano},$^1$   Yoshihiro  {\sc Nitta},$^1$     and Yoshiaki {\sc \=Ono}$^2$}
\inst{$^1$Department of Physics Engineering, Mie University, Tsu, Mie 514-8507, Japan \\
$^2$Department of Physics, Niigata University, Ikarashi,  Niigata, 950-2181, Japan}


\abst{
Using first-principles calculations, we  examine the  transition temperature  $T_{\rm c}$ of  superconductivity in sodium tungsten bronze ( Na$_x$WO$_3$, where $x$ is  equal to or less than unity ).
Although  $T_{\rm c}$ is  relatively low  $T_{\rm c}( \simk 3 {\rm K})$,
it is interesting that its characteristic exponential dependence on $x$   has been  experimentally observed at $0.2 \simk x \simk 0.4$.
On the basis of the McMillan equation for $T_{\rm c}$ including the effect of  plasmons, we succeed in  reproducing the absolute values of $T_{\rm c}$ and its $x$ dependence. 
We also find that the  plasmon effect is crucial for the estimation of $T_{\rm c}$ as well as phonons.
Since the calculated $T_{\rm c}$ may not  exceed $\sim 20$ K  even for $x \simk 0.1$,  the  superconductivity at a  low $T_{\rm c}$ can be interpreted by the usual phonon  mechanism, including the  plasmon effect.
On the other hand, a high $T_{\rm c}$  up to about 90 K, which is found on the surface of a Na$_x$WO$_3$ system at $x\sim 0.05$ by recent experiments,  cannot be explained by our results.
This discrepancy suggests that another  mechanism is required to clarify the nature of the high-$T_{\rm c}$ superconductivity of Na$_x$WO$_3$.
}

\maketitle
Sodium tungsten bronze ( Na$_x$WO$_3$ ) and related materials [ A$_x$WO$_3$(A= K, Rb, Cs,..) ]  have been studied for a long time as  typical oxide superconductors, where the concentration  $x$ is  equal to or less than unity.\cite{Shanks-1974,Bloom-1976,Skokan-1979,Cadvyell-1981,Leitus-2002,Brusetti-2002,Raj-2007,Brusetti-2007,Haldolaarachchige-2014}  
The transition temperature $T_{\rm c}$ of superconductivity has been known to  increase with  decreasing $x$  and to be up to about 3 K for Na$_x$WO$_3$  ( $\sim$ 7 K for Cs$_x$WO$_3$)  at $x \sim 0.2$.
It is interesting that the $x$ dependence of $T_{\rm c}$ is experimentally
 given by $T_{\rm c} \simeq {\rm A} \exp(-{\rm B}x)$ for $0.2 \simk x \simk 0.4$ for Na$_x$WO$_3$\cite{Shanks-1974}, where A and B are constants.
Since $T_{\rm c}$ is relatively low, the origin of this superconductivity has been explained by the usual phonon  mechanism.\cite{Nagi-1976}

However,  recent experiments  on the  surface of a Na$_x$WO$_3$ system showed that  high-temperature superconductivity (HTS) at about 90K or higher is achieved   at $x \sim 0.05$  for Na$_x$WO$_3$.\cite{Levi-2000,Aliev-2008}
Furthermore, another experiment demonstrated that  H$_x$WO$_3$ also shows HTS at about 120K.\cite{Reich-2009}  
  
The result of scanning tunneling microscopy\cite{Levi-2000} indicates that the $I-V$ characteristic behavior is well fitted to the BCS theoretical curve. 
It seems to suggest that  the symmetry of a Cooper pair may be described by the  $s$-wave and it  does not contradict the phonon-mediated superconductivity.
These results are extremely surprising because  the values of $T_{\rm c}$ are comparable to those   of cuprate superconductors.
If the HTS of Na$_x$WO$_3$ and/or H$_x$WO$_3$ are concrete,  it is very interesting to clarify whether the   mechanism underlying this superconductivity  is the conventional one or not.

In this study,  we  examine the electronic state  of Na$_x$WO$_3$  by   first-principles calculations  using  'Quantum ESPRESSO', which is an integrated software of Open-Source computer codes for electronic-structure calculations.\cite{QE} 
Assuming the phonon-mediated  superconductivity, we  estimate  $T_{\rm c}$ as  a function of  $x$  on the basis of the   McMillan formulation\cite{McMillan,Allen,Morel}  including the effect of  plasmons.\cite{Takada-1978,Bill-2002,Akashi-2013,Akashi-2014,Sano-2019}
Since  $T_{\rm c}$ can be obtained with high accuracy by first-principles calculations, we may obtain  a clue  distinguishing whether the observed HTS can be explained by the conventional mechanism or not.

%
%
\begin{figure}[thb]
\begin{center}
\includegraphics[width=0.65 \linewidth]{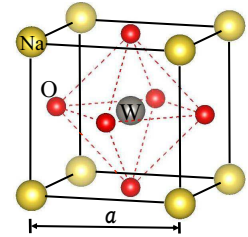}
\end{center}
\caption{(Color online)
Unit cell  structure of   sodium tungsten bronze,  NaWO$_3$, where  $a$ is the length of one side of the  cell.
Inside the cube,  one tungsten atom and  six oxygen atoms connected by  dotted lines represent the octahedral WO$_6$.
}
\label{model}
\end{figure}

\begin{figure}[thb]
\begin{center}
\includegraphics[width=0.7 \linewidth]{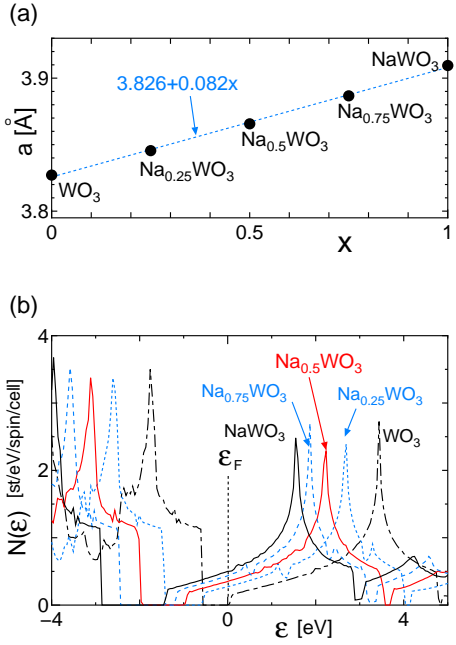}
\end{center}
\caption{(Color online)
  (a) Calculated equilibrium  values of $a$  as a function of $x$, where $x$ is composition ratio of Na.
(b) Density of states of  Na$_x$WO$_3$ calculated for the supercell, where $x=0$ (doted-dashed line), 0.25 (doted line), 0.5 (solid line), 0.75(broken line), and 1.0(solid line), where  the values of DOS ($N(\epsilon)$) are normalized  by a single unit cell.
}
\label{v-a}
\end{figure}

In Fig.\ref{model}, we show a schematic structure of the unit cell of Na$_x$WO$_3$   for $x=1$.
It  has a cubic perovskite  structure, where the length of one side of the cube is $a$.
Inside the cubic unit cell,  one tungsten atom and six oxygen atoms form an octahedron, which we will call  octahedral WO$_6$.
First, we calculate the $x$ dependence of  $a$ using a supercell  consisting of  four unit cells.  It contains zero, one, two,  three, and four Na atoms. 
Here,  we  use an $8 \times 8 \times 8$  Monkhorst$-$Pack (MP)  grid for  electronic Brillouin zone integration. 
Using a structural relaxation method, we determine the volume of the supercell and estimate the value  corresponding to $a$  as the cubic root of one quarter of its volume.  
It is experimentally  known that  in a system of  Na$_x$WO$_3$, the unit cell size  $a$ decreases with decreasing  $x$.\cite{Brown-1954}
In Fig. \ref{v-a}(a), we show $a$ as a function of $x$.
It indicates that that $x$ dependence of $a$  is given by $a \simeq 3.826+0.082x$ [$\AA$].
The result well reproduces that obtained by  experiments as
$a=3.7845+0.0820x$ [$\AA$].\cite{Brown-1954}

 In  Fig. \ref{v-a}(b),  we also show the density of states (DOS) for the same supercells.
For $x=0$, the energy gap  appears at the Fermi energy and the system becomes an insulator.
On the other hand, the value of DOS at the Fermi energy  is finite and the system  is metallic for $x>0$. 
We confirmed that the DOS of one Na atom is  less than $10^{-3}$ [st/eV/spin/cell] near the Fermi energy, and it is  completely  negligible for  $x \ge 0$.\cite{Hjelm-1996} 
Furthermore, the energy dependences of DOS are very similar except for the position of the Fermi energy, as shown in Fig. \ref{v-a}(b). 
Therefore, the role of the Na atom is  considered to only provide an electron into the conduction  band  of the system of WO$_3$.
These results suggest that the rigid-band picture well stands, as shown by Raj et al.\cite{Raj-2007}. 
Therefore,  to calculate in the case of an arbitrary $x$, we  adopt the rigid-band approximation (RBA).\cite{Subedi}  This approximate method   introduces a  fictitious carrier in the target system by assuming  a rigid-band. It changes only the Fermi energy of the system according to the carrier density.  Although it is a simple approximation, it allows us to systematically analyze the   electronic state  of  Na$_x$WO$_3$ for $x \le 1$.\cite{Subedi,Sano-2019}   
Hereafter, we denote the  result obtained by adopting RBA for the WO$_3$ system as  WO$_3^{\rm RBA-}$  at an arbitrary $x$, whereas that for  NaWO$_3$  as NaWO$_3^{\rm RBA+}$.
Here, the notation of ${\rm RBA-}$ means the case of electron doping in WO$_3$, and  ${\rm RBA+}$ means that of hole doping in NaWO$_3$.
%
%

If the superconductivity of Na$_x$WO$_3$  is caused by  phonon-mediated attraction,
  $T_{\rm c}$ can be  estimated  by the following  McMillan equation: 
\begin{align} 
T_{\rm  c} \simeq \frac{\omega_{\rm  log}}{1.2}\exp(-\frac{1.04(1+\lambda)}{\lambda-\mu^*(1+0.62\lambda) }),
\label{M_Tc}
\end{align} 
where $\omega_{\rm log}$ is the logarithmic average  frequency, which means a characteristic phonon  frequency of the system, $\lambda$ is the electron-phonon coupling constant, and $\mu^*$ is the  Coulomb pseudopotential, which  is usually treated as a constant of about 0.1 for  metals.\cite{Morel}
\begin{figure}[thb]
\begin{center}
\includegraphics[width=0.7 \linewidth]{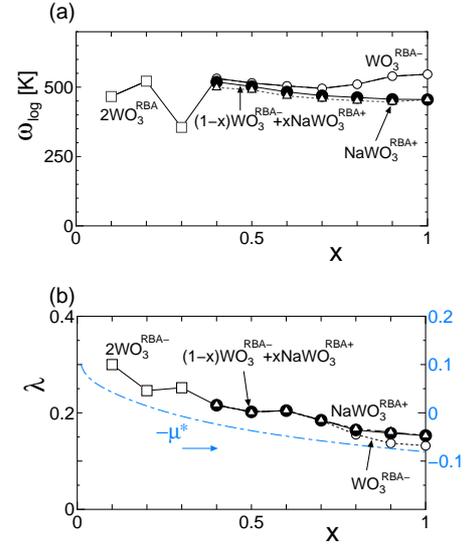}
\end{center}
\caption{(Color online)
 (a) Logarithmic average  frequency $\omega_{\rm log}$ as a function of $x$ for WO$_3^{\rm RBA-}$ (open circles),  Na WO$_3^{\rm RBA-}$ (open triangles), and    arithmetic average of both (solid circles), where  the  MP  grid used is  $24 \times 24 \times 24$.  
It is noted that open triangles are very close to solid circles, and the former seem to appear   above the latter.
Open squares are the result of the supercell made of two unit cells, 2WO$_3^{\rm RBA-}$,  where the  MP  grid used is  $8 \times 8 \times 8$.  
(b)  Electron-phonon coupling constant $\lambda$  as a function of $x$.
The dotted-dashed line represents $-\mu^*$ as a function of $x$.
}
\label{lambda}
\end{figure}
However, $\mu^*$ is dependent on  the electron density of the system, $n$  for  $n \simk 10^{22}$ cm$^{-3}$  owing to the effect of  plasmons.\cite{Takada-1978,Sano-2019} 
When  $n$ decreases,   $\mu^*$ also decreases and it even becomes negative at $n \sim 3 \times 10^{21}$ cm$^{-3}$.\cite{Sano-2019,Klein-2007}
In the system of Na$_x$WO$_3$,  $n$ is given by $ \simeq 1.7 \times 10^{22}  {\rm cm^{-3}}$ for $x=1$, and   $\mu^*$   is expected to be considerably smaller than 0.1  in the region of $x \simk 0.5$.\cite{Nagi-1976,Sano-2019}  
%

In this work, we adopt a simple phenomenological relation,
\begin{align} 
\mu^* \simeq 0.05\{\log_{10}(x)+\log_{10}(1+40x^2)\},
\label{pheno}
\end{align} 
which well reproduces the $x$ dependence of $\mu^*$ obtained in the previous work for $0.01 \simk x \simk 0.5$.\cite{Sano-2019}  
This relation allows us to estimate T$_{\rm c}$ including the  plasmon effect  through  the McMillan equation (\ref{M_Tc}).
Using the  first-principles calculations, we  can estimate  two parameters  $\omega_{\rm log}$ and  $\lambda$ precisely. 
Combined with the above  $\mu^*$, we can obtain the concrete value of $T_{\rm  c}$  by this formulation.

To calculate $x$ dependences of $\omega_{\rm log}$ and  $\lambda$ by RBA, we use  two independent results, WO$_3^{\rm RBA-}$ and NaWO$_3^{\rm RBA+}$ in the case of $x \ge 0.4$. 
We assume that the electronic properties of Na$_x$WO$_3$ corresponding to an arbitrary $x$ are approximately obtained  by prorating   both results, such as  $(1-x)$WO$_3^{\rm RBA-}+x$NaWO$_3^{\rm RBA+}$.
Here, we use the $24 \times 24 \times 24$  MP  grid for the electronic state and the $4 \times 4 \times 4$ grid for the phonon calculation.\cite{smearing1}  
The unit cell size  is chosen to be the same as that  given in  Fig.1 for both WO$_3$ and NaWO$_3$ unit cells.
This approximation method may  correspond to  mean field approximation, which neglects the effect of randomness produced by Na atoms.

In Fig. \ref{lambda}(a), we show $\omega_{\rm log}$ as a function of $x$.   $\omega_{\rm log}$ seems to be  almost constant and  is roughly given by 500 K. 
The difference in  $\omega_{\rm log}$ between  WO$_3^{\rm RBA-}$   and NaWO$_3^{\rm RBA+}$  is small. 
Therefore,  the  prorating method may be a good approximation to consider the system of  Na$_x$WO$_3$.
In Fig. \ref{lambda}(a), the values of  $\lambda$ and  $-\mu^*$ as  functions of $x$ are given. 
The result   indicates that the difference in  $\lambda$ between  WO$_3^{\rm RBA-}$   and NaWO$_3^{\rm RBA+}$  is very small.
 $\lambda$  increase with decreasing  $x$, which is roughly  similar to the result obtained by  Mascello et al. \cite{Mascello-2020}
The figure  also shows that  $-\mu^*$   increases with decreasing  $x$,  and  its rate of change  is  almost  the same as that of $\lambda$.
This  means that the  plasmon effect described by $\mu^*$  plays as important role in the superconductivity of Na$_x$WO$_3$  as  the phonon-mediated attraction  described by  $\lambda$.  

%
%
%
\begin{figure}[hbt]
\begin{center}
\includegraphics[width=0.7 \linewidth]{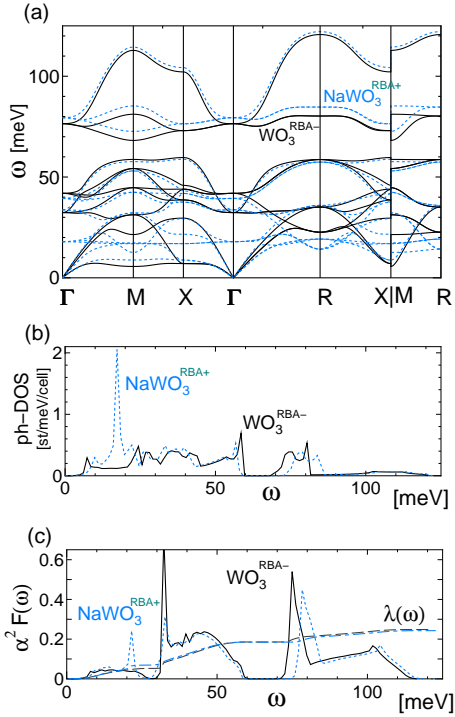}
\end{center}
\caption{(Color online) 
(a)  Phonon dispersions of  WO$_3^{\rm RBA-}$ (solid  lines)  and NaWO$_3^{\rm RBA+}$ (dotted lines) for $x=0.4$. 
(b) Phonon DOS, $F(\omega)$ of  WO$_3^{\rm RBA-}$ (solid  line)  and NaWO$_3^{\rm RBA+}$(dotted line) for $x=0.4$. 
(c) Electron-phonon spectral function $\alpha^2 F(\omega)$ for $x=0.4$, where 
the dashed (WO$_3^{\rm RBA-}$)  and  dotted-dashed (NaWO$_3^{\rm RBA+}$) lines stand for $\lambda(\omega)$ defined as 
 $\lambda (\omega)=2\int_0^\omega \alpha^2 F(\omega')/\omega' {\rm d}\omega'$. }
\label{a2F}
\end{figure}

For $x\simk 0.4$, the  symmetry of a unit cell is not  cubic, but  tetragonal,  orthorhombic,  or triclinic depending on the value of $x$.\cite{Shanks-1974,Raj-2007}
In this case,  the shape of a unit  cell changes and the arrangement of  octahedral  WO$_6$  becomes  complicated.\cite{Shanks-1974,Raj-2007} 
However,  the distortion of  octahedra WO$_6$ is small and its  structure is  close to that when it is inside a cube.\cite{Wijs-1999, Walkingshaw-2004} 
Therefore, we consider only  the contribution of part of WO$_3$, that is,   octahedral WO$_6$,  and neglect the  effects  of Na atoms except  for the effect of supplying  carriers  to the system.
From the result  for $x \geq 0.4$  shown in Fig. \ref{lambda},  the effect of  Na atoms is expected to be small  in the region of  $x \simk 0.4$. Therefore,  this approximated treatment may be allowed for the estimation of $T_{\rm c}$.
In fact,  DOS of  Cs$_x$WO$_3$\cite{Ingham-2005}, which has a hexagonal structure, is  similar  to that of Na$_x$WO$_3$ near the Fermi energy despite  the difference in crystal structure. 
Furthermore,   the absolute value and   the $x$-dependence  of $T_{\rm c}$ obtained in experiments\cite{Shanks-1974, Skokan-1979} seem to be also similar for both systems. 

To examine the effect of Na atoms in more detail, we calculate  phonon dispersion, phonon DOS,  and the electron-phonon coupling   function $\alpha^2 F(\omega)$ for WO$_3^{\rm RBA-}$ and NaWO$_3^{\rm RBA+}$  at $x=0.4$.\cite{smearing2} 
Here, an integral  of $\alpha^2 F(\omega)/\omega$  is related  to $\lambda$ as $\lambda (\omega)=2\int_0^\omega \alpha^2 F(\omega')/\omega' {\rm d}\omega'$.
Figure \ref{a2F}(a)  indicates that the phonon dispersions of  both WO$_3^{\rm RBA-}$ and NaWO$_3^{\rm RBA+}$ are very similar, except for the almost flat mode of  NaWO$_3^{\rm RBA+}$ at $\omega \simeq 18  {\rm meV}$.
As shown in  Fig. \ref{a2F}(b),  this mode  appears as  a sharp peak at  $ \simeq18 {\rm meV}$ in the phonon DOS of  NaWO$_3^{\rm RBA+}$, whereas there is no corresponding peak in that of  WO$_3^{\rm RBA-}$. 
It can be interpreted as an oscillation mode of  Na atoms in the cell of NaWO$_3$. Since there seems to be no dispersion,  the vibration of Na atoms behaves as an Einstein phonon with a frequency  $ \simeq18 {\rm meV}$.
It also indicates that  Na atoms are isolated in the crystal, and  phonons of Na have little  effect on those of other atoms.
A similar situation has been noted on  the Rb$_x$WO$_3$\cite{Brusetti-2007}  and Cs$_x$WO$_3$\cite{Pellegrini-2019}  systems. 

Figure \ref{a2F}(c) shows that  the main contribution to  $\lambda$ comes from part of WO$_3$ and the contribution from Na  may be negligible. This result is consistent with the electron DOS of Na atoms being almost zero at the Fermi energy\cite{Hjelm-1996}
This figure also shows that the contribution of acoustic phonon modes near the $\Gamma$-point ($\omega \simk 30  {\rm meV}$)  is not large, and the main part of $\lambda$ comes from the optical modes of WO$_3$ for $\omega \simj  30  {\rm meV}$, by considering it against the result showing in Fig. \ref{a2F}(a).
If  long-wavelength  modes such as  acoustic phonons  are not essential  for the evaluation  of  $\omega_{\rm log}$ and $\lambda$,   the  arrangement of octahedral WO$_6$  is expected to be not important, and  the above-mentioned simplifications omitting the effects of Na atoms  may be justified for the estimation of $T_{\rm c}$.

Since the tungsten atoms are staggered in adjacent unit cells in the case  of  $x \leq 0.3$\cite{Wijs-1999, Walkingshaw-2004}, we use a supercell made of two unit cells of WO$_3$ as a new unit cell,  2WO$_3$,  to calculate $\omega_{\rm log}$  and $\lambda$. 
The  structure of 2WO$_3$ and atomic positions are determined by  the structural relaxation  within RBA. 
For $x=0.2$ and 0.3,  we find that the suitable structure of WO$_3$ is  tetragonal, whereas it is orthorhombic for $x=0.1$ owing to the subsequent structural transition.\cite{Raj-2007}


\begin{figure}[th]
\begin{center}
\includegraphics[width=0.8 \linewidth]{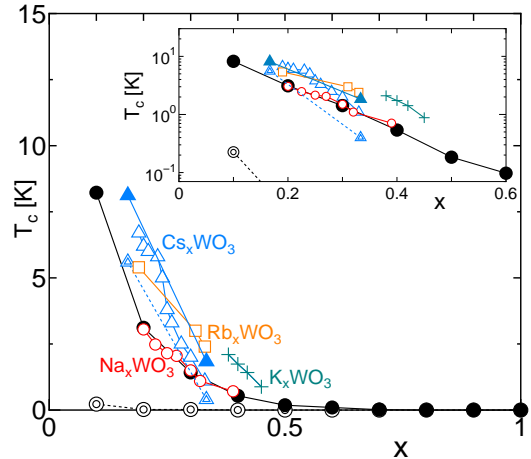}
\end{center}
\caption{(Color online) Calculated $T_{\rm c}$ as a function of  $x$ with the results of experiments. 
Here,  solid circles  represent our calculated $T_{\rm c}$ and double circles with the dotted line represent that excluding the plasmon effect.  Open circles indicate  experimental $T_{\rm c}$ of Na$_x$WO$_3$\cite{Shanks-1974}, open triangles represent  that  of Cs$_x$WO$_3$.\cite{Skokan-1979}  Crosses indicate  K$_x$WO$_3$\cite{Haldolaarachchige-2014}, and   open squares indicate  Rb$_x$WO$_3$\cite{Brusetti-2002}. 
Double  triangles with the dotted line indicate the theoretical estimation of $T_{\rm c}$ for Cs$_x$WO$_3$ obtained by  Pellegrini et al.\cite{Pellegrini-2019} and solid triangles indicate that added to the plasmon effect.
The inset shows a semi-log graph for $T_{\rm c}$ as a function of  $x$ }
\label{Tc}
\end{figure}
In Fig. \ref{Tc}, we show the calculated $T_{\rm c}$ as a function of  $x$ with several  results of experiments and that of theoretical result for Cs$_x$WO$_3$.\cite{Pellegrini-2019}
Here,  to clarify the effect of plasmons, we show the theoretical results  with  and without  the plasmon effect.
The former (solid circles) is in good agreement with the experimental results of Na$_x$WO$_3 $  (open circles), whereas values of the latter\cite{noplasmon}  (double circles) are close to 0K, indicating that they are too low.
These results suggest that the plasmon effect is  important to explain the superconductivity of  Na$_x$WO$_3$ as well as the phonon mechanism.

The  theoretical estimation obtained by Pellegrini et al.\cite{Pellegrini-2019}  for Cs$_x$WO$_3$ (double  triangles) does not include the plasmon effect. 
Therefore, if we add the effect to their result, the obtained values of $T_{\rm c}$ (solid  triangles)  increase and seem to be in good  agreement   with the result of  the experiment (open triangles). Here, to estimate $T_{\rm c}$, we combined the values  of $\lambda$ obtained by  Pellegrini et al.\cite{Pellegrini-2019} and the phenomenological relation (\ref{pheno}) for $\mu^*$, and $\omega_{\rm log}$ is assumed to be  200 K for simplicity.
This result also indicates the importance of the plasmon effect for superconductivity.\cite{Takada-1978,Bill-2002,Akashi-2013,Sano-2019}  
It is interesting that   absolute values  and the $x$ dependence of  $T_{\rm c}$ for   Na$_x$WO$_3$ seem to be  similar  to those for Cs$_x$WO$_3$ and other materials despite  differences in the crystal structures.\cite{Ingham-2005,Pellegrini-2019}
This suggests that the  octahedra  common in these materials are mainly responsible for their superconductivity.

In an actual Na$_x$WO$_3$ system, it becomes an insulator when $x$ is smaller than about 0.2 and its superconductivity is not observed.
Because the  structural transition occurs and the crystal symmetry of the system  changes,   the effect of the Anderson localization may be  enhanced, which causes the metal-insulator transition observed at $x \sim 0.2$.\cite{Raj-2007} 
Note that our result is obtained from a perfect crystal without randomness, rather than realistic crystals with randomness.

The inset in Fig. \ref{Tc} shows a semi-log plot of $T_{\rm c}$. It indicates that  the $x$-dependence of $T_{\rm c}$ seems to be given by
$T_{\rm c} \simeq {\rm A} \exp(-{\rm B}x)$ for  $0.1 \simk x \simk 0.6$, where  A and B are constants.
We find these values to be A=20.7 [K]  and B=9.3 by  fitting to the data of  Na$_x$WO$_3$.
This result gives a theoretical base  for understanding  the exponential behavior of  $T_{\rm c}$ obtained by the experiment for $0.2 \simk x \simk 0.4$.\cite{Shanks-1974}
The inset also indicates that  $x$-dependences are almost the same regardless of the material except  for the theoretical  results which have no  plasmon effect. 
By  including  the plasmon effect to the theory, we can  obtain good agreement with the experiments.

Finally, we discuss the relationship between our result and the HTS of Na$_x$WO$_3$.
If we use the above fitting equation for $T_{\rm c}$, it becomes about 13 K at  $x=0.05$. The result is clearly  inconsistent with the result of the experiment  on  HTS.
Of  course,  HTS is observed on the surface and we should pay attention to the dimensionality of the system. 
The surface superconductivity has  already been examined by the first-principles calculations, although the result  is limited   to  hole doped  diamond systems.\cite{Nakamura-2013,Sano-2017}
Results indicate that  the  electronic state forms a characteristic  bound state on the surface. However,  the values of  $\omega_{\rm log}$  and $\lambda$ are   comparable to those  of the bulk,  and no special mechanism  was found to  significantly increase these values.\cite{SrTiO3}

On the other hand, the absolute value of  $\mu^*$ increases with decreasing  dimensionality of the system.
Using the result of  Takada\cite{Takada-1978}, we find that  the absolute value of  $\mu^*$ increases by about 0.05 at $x=0.05$ for  a pure two-dimensional system.\cite{2d}
If we assume   $\omega_{\rm log}=500$ K  and $\lambda=0.3$\cite{x005}, which are typical values for a bulk system as shown in Fig. \ref{lambda}, $T_{\rm c}$ including  the enhancement of $\mu^*$ becomes about 20K.  However, this value  is far from $\sim 90$ K of HTS. 
Therefore,  it is difficult to explain the HTS of Na$_x$WO$_3$ by the conventional phonon mechanism including the plasmon effect,  and some new  mechanisms such as orbital fluctuation\cite{Sekikawa-2020}  may be required.

In summary,  we have investigate the $x$ dependence of $T_{\rm c}$ in sodium tungsten bronze (Na$_x$WO$_3$)  first-principles calculations.
We find that the superconductivity  is dominated by a part of  WO$_3$, and the role of  Na atoms is almost limited to providing  carriers to the conduction band of  the system. 
Combined with the McMillan equation and the phenomenological relation of  $\mu^*$ including the plasmon effect,  we show that  $T_{\rm c}$ is given as   $T_{\rm c} \simeq 20.7 \exp(-9.3x)$ for $0.1 \simk x \simk 0.6$, which is consistent with the result obtained  by the experiment for $0.2 \simk x \simk 0.4$. We have found that  plasmons are as important  as phonons for the superconductivity of Na$_x$WO$_3$.
Although $T_{\rm c}$ increases with decreasing $x$, it may not exceed $\sim 20$K  for small $x$.  
This result is  inconsistent with HTS up to about 90 K  at $x \sim 0.05$ observed in the recent experiment for the surface.
This discrepancy may require some new mechanism  to explain the HTS of Na$_x$WO$_3$ beyond the conventional  phonon mechanism.

In this work, we used the Gaussian smearing method to obtain 
 $\lambda$, $\omega_{\rm log}$, and  $\alpha^2 F(\omega)$.
The possible errors of these values are not large.\cite{smearing1}
However, more efficient method introduced by Koretsune and Arita\cite{Koretsune}
 may give  more accurate results for these values.
In addition,  we used the phenomenological relation of $\mu^*$ and the McMillan equation to calculate $T_{\rm c}$, but it is insufficient  from a theoretical  viewpoint.
More microscopic theories such as superconducting density functional theory including the plasmon effect will be required.\cite{Akashi-2013,Akashi-2014}
We would address these problems  in our  future study.


\begin{acknowledgment}
This work was supported by JPSJ KAKENHI Grant Numbers  17K05539 and 19K03716. 
The authors thank Takuya Sekikawa, Rai Watabe, Jun Ishizuka, and Kouki Hara for valuable discussions. 
\end{acknowledgment}


\begin{thebibliography}{99}
\bibitem{smearing1}  
To  estimate $\lambda$ and $\omega_{\rm log}$,  we used the   Gaussian smearing width as 0.015 0.02, 0.025, 0.03, 0.035, and 0.04 [Ry], and
 calculated average values  at these smearing values.
In a typical case of  $x=0.4$,  standard deviations of 
 $\lambda$ and $\omega_{\rm log}$  were  about 0.03 and  about 2 [K], respectively.
\bibitem{smearing2}  
Gaussian smearing width 0.025 [Ry] was used for the electron-phonon coupling function $\alpha^2 F(\omega)$.
\bibitem{noplasmon}
To eliminate the plasmon effect, we  simply fix $\mu^*$ at 0.1 and calculate $T_{\rm c}$.
\end{thebibliography}
\end{document}